\documentclass{article}
\usepackage{spconf,amsmath,graphicx}
\usepackage{comment}
\usepackage{here}

\newcommand{\fnote}[1]{}

\newcommand{\knote}[1]{{\color{red} \bf #1 \color{black}}}

\newcommand{\bnote}[1]{{\color{red} #1 \color{black}}}
\newcommand{\mnote}[1]{}
\newcommand{\onote}[1]{\color{blue} #1 \color{black}}
\newcommand{\snote}[1]{}
\newcommand{\kcut}[1]{}
\newcommand{\ocut}[1]{}
\newcommand{\scut}[1]{}
\newcommand{\jptext}[1]{}

\ifx\pdfoutput\undefined
\else
 \usepackage{CJKutf8}
 \renewcommand{\fnote}[1]{}
 \renewcommand{\knote}[1]{}
 \renewcommand{\mnote}[1]{}
 \renewcommand{\bnote}[1]{}
 \renewcommand{\onote}[1]{}
 \renewcommand{\jptext}[1]{}
\fi

\newcommand{\bm}[1]{\mbox{\boldmath{$#1$}}}
\ifx\etal\undefined\newcommand{\etal}{{\it et al. }}\fi
\ifx\ie\undefined\newcommand{\ie}{{\it i.e.}}\fi
\ifx\eg\undefined\newcommand{\eg}{{\it e.g.}}\fi

\newcommand{\figref}[1]{{Fig.~\ref{fig:#1}}}

\newcommand{\subsecref}[1]{Sec.~\ref{ssec:#1}}

\title{\vspace{-0.2cm}
Underwater stereo using REFRACTION-FREE image\\ synthesized from LIGHT FIELD CAMERA
\vspace{-0.2cm}}
\name{Kazuto Ichimaru \qquad Hiroshi Kawasaki
\vspace{-0.2cm}}
\address{Kyushu University, Japan
\vspace{-0.2cm}}
\begin{document}
%\ninept
%
\maketitle
\begin{abstract}
There is a strong demand on capturing underwater scenes without distortions caused 
    by refraction.
Since a light field camera can capture several light rays at each point of an image plane from 
    various directions, if geometrically correct 
    rays are chosen, %at each point, 
it is possible to synthesize a refraction-free image.
In this paper, we propose a novel technique to efficiently select such rays to synthesize a refraction-free 
    image from an underwater image captured by a light field camera.
%Some conventional methods focused on making refraction-free image by adjusting focal length and lens distortion approximation of refraction effect.
%However, they produce approximation error when the setup is changed from specific condition.
%Using light field camera, it is possible to make geometrically correct refraction-free images without such approximation error.
In addition, we propose a stereo technique to reconstruct 3D shapes 
using a pair of our refraction-free images, which are central projection.
%, which are synthesized by our technique.
%common passive stereo techniques developed for ordinary cameras.
In the experiment, we captured several underwater scenes by two light field 
    cameras, synthesized refraction free images and 
    applied stereo technique to reconstruct 3D shapes. % of the scene.
%on refrection free images synthesized by our algorithm. 
The results are compared with previous techniques which are based 
    on approximation, showing the strength of our method.
%We also describe a scheme of passive stereo using proposed algorithm as well as evaluation compared to conventional algorithm.
%In the experiment, we show the superiority of proposed algorithm in terms of computation cost and accuracy of produced refraction-free images.

\end{abstract}

\vspace{-0.5cm}
\section{Introduction}
\vspace{-0.3cm}
\label{sec:intro}
Underwater scene acquisition is an important research topic for various areas, such 
as underwater construction, marine biology, swimming analysis to name a few.
For those purposes, 3D shape reconstruction is most important, %a key to capture underwater environment, 
and thus, many techniques have been researched and developed.
In terms of 3D reconstruction, passive stereo using two RGB cameras is commonly 
used in the air, 
because 
of its simplicity and stability.
Since ordinary cameras are usually perspective, which means central 
projection, most stereo techniques also assume the central projection, 
which enables 1) efficient correspondence search using epipolar constraint, 
2) linear solution on calibration, and 3) shape reconstruction by triangulation.
%linear light path constraint and epipolar constraint,
%which enables to derive linear solution and makes the computation tractable and much simpler.
However, underwater images are not central projection because of refraction, 
%between water and waterproof housing, 
and thus, it is  difficult to apply common 
stereo techniques to underwater scene.

%Non-linear light path formulations are often intractable and also epipolar line becomes non-linear.

To overcome the problem, analytical solution to estimate light path of 
refraction by solving high order simultaneous equations is proposed~\cite{Agrawal:CVPR2012}.
However, solving such equations requires high computational cost and it 
also remains ambiguities.
%light path tracing, including efficient ways to obtain initial values or bundle adjustment equations~\cite{Agrawal:CVPR2012}.
%However, as shown in the paper, analytical forward projection of non-central camera model costs heavy computational time.
On the other hand, approximation based technique to synthesize a refraction-free 
image is proposed~\cite{Ferreira:PRIA2005}.
Although the synthesized refraction-free image can be treated as central 
projection at predefined depth,
%if it has no refraction and enables lightweight computation,
approximation error increases when object depth becomes far  from the depth.
%, or irregular refraction due to slanted housing surface.
%Lack of original (\ie, central projection) light ray is the main reason of impossibility to make geometrically correct refraction-free image.

Recently, light field imaging draw a wide attention for its potential on
post-focusing, single view stereo and so on. %, and recovering occluded region.
Since a light field data is a collection of multiple light rays from various 
directions including non-central rays at each point on image plane, 
refraction-free image can be created by choosing a certain light ray from a 
bundle of rays at each point.
In the paper, we propose a technique to find geometrically correct light rays 
from a light field data to synthesize a refraction-free image.
Our method is implemented as a pixel warping on a captured light field 
image, 
%a pre-compution of
thus, calculation time is comparable to approximation based approach.
%much faster than analytical light path tracing.
In addition, we propose a stereo technique using a pair of refraction-free 
images to reconstruct 3D shapes of underwater environment. 
In the experiment, we captured several underwater scenes by two light field cameras and 
    conducted stereo technique to reconstruct 3D shapes.
The results are compared with previous techniques, which are based 
    on approximation model, successfully showing the strength of our method.

\vspace{-0.35cm}
\section{Related work}
\vspace{-0.3cm}
\label{sec:related}
There are generally two approaches to handle refraction between the water and the air.
%in underwater computer vision; 
The one is geometric approach \cite{Agrawal:CVPR2012,Sedlazeck:ICCV2013,Kawahara:ICCVW2013} 
and the other is approximation-based approach \cite{Ferreira:PRIA2005,Bleier:ISPRS2017,Kawasaki:WACV2012,Ichimaru:3DV2018}.
Geometric approach considers physical model of refraction to trace light rays, 
and applies forward / backward projection to render 2D images from 3D objects.
%triangulation with analytical 
In those methods, several parameters are necessary to be calibrated, such as 
refractive index, distance between camera and the water and normal of refractive interface.
Agrawal \etal introduced a polynomial formulation for the model and efficient 
initial value computation method as well as analytical forward projection by 4-th order equation~\cite{Agrawal:CVPR2012}.
%They also described analytical forward projection requires to solve 4-th order equation under two mediums with different refractive indices,
%and 12-th order equation under three mediums.
Sedlazeck and Koch proposed underwater SfM to directly recover 
3D shapes without explicit calibration using efficient energy function for bundle adjustment by considering virtual camera~\cite{Sedlazeck:ICCV2013}.
Kawahara \etal proposed a pixel-wise varifocal camera model~\cite{Kawahara:ICCVW2013}.
Note that most of the geometric approaches introduce special models to represent 
refraction effects, which is not a central projection model, and thus, general 
stereo method cannot be applied.
%3D reconstruction methods.
%Moreover, heavy computational cost and instability reduce practicality.

Approximation-based approach usually converts an original image to a 
central projection image.
% introduces virtual focal length to make refracted light path as linear as possible, 
%and absorbs remaining error by approximation to lens distortion.
Ferreira \etal proposed an approximation-based technique to make refraction-free 
image by applying lens distortion model, and showed some results on stereo vision~\cite{Ferreira:PRIA2005}.
%Bleier and N\"{u}chter used a cross line laser projector for 3D scan~\cite{Bleier:ISPRS2017},
%and a camera~\cite{Bleier:ISPRS2017},
%Kawasaki \etal used a diffractive optical element to achieve one-shot dense reconstruction~\cite{Kawasaki:WACV2012}
%and Ichimaru \etal extended the technique by using a stereo camera 
%pair~\cite{Ichimaru:3DV2018} based on approximation-based approach.
Bleier and N\"{u}chter~\cite{Bleier:ISPRS2017}, Kawasaki 
\etal\cite{Kawasaki:WACV2012}, and Ichimaru \etal\cite{Ichimaru:3DV2018} applied 
active method based on approximation-based approach.
Since those techniques assume specific depth, %for approximation, 
severe 
distortion occurs if the actual depth is far from the predefined depth. 
%predefined depth. 
%work well in specific condition, \ie, depth of the target object is close enough to calibrated depth within limited range, as insisted in \cite{Ichimaru:3DV2018}.

Recently, light field imaging technique draw a wide attention.
%is widely used in the field of image 
%processing, shape reconstruction, projection mapping, and so on.
%Applications of light field imaging is not limited to light field camera, but also light field projector or light field probes.
In image processing task, Lu \etal used light field images with CNN for depth map restoration to cope with turbidity of water~\cite{Lu:AIR2010},
and Li \etal used them for reflection removal~\cite{Li:TIP2018}.
%In general 3D reconstruction task, 
Jeon \etal proposed an accurate depth map estimation method using light field camera~\cite{Jeon:CVPR2015}.
Zhang \etal proposed plenoptic SfM technique~\cite{Zhang:ICCV2017}.
Kutulakos and Steger achieved reconstruction of static transparent objects by light field imaging~\cite{Kutulakos:IJCV2008}.
%and in transparent object reconstruction task, 
Wetzstein \etal used light field probes for normal and shape reconstruction of dynamic transparent objects~\cite{Wetzstein:ICCV2011}.
Skinner and Roberson used light field camera for single-view underwater 3D reconstruction, but they ignored refraction~\cite{Skinner:IROS2016}.
Among wide variety of light field research,
synthesis of refraction-free image has not been studied yet.

% ライトフィールド系の関連研究に何を入れるかがまだ確定していない
\begin{comment}
Image processing
	Underwater Light Field Depth Map Restoration Using Deep Convolutional Neural Fields (H. Lu et al. Artificial Intelligence and Robotics 2018)
	Robust reflection removal based on light field imaging (T. Li et al. Transactions on Image Processing2018)

General 3D reconstruction
	Accurate Depth Map Estimation from a Lenslet Light Field Camera (H.-G. Jeon et al. CVPR2015)
	Ray Space Features for Plenoptic Structure-from-Motion (Y. Zhang et al. ICCV2017)

Transparent 3D reconstruction
	A Theory of Refractive and Specular 3D Shape by Light-Path Triangulation (Kutulakos and Steger. IJCV2008)
	Refractive Shape from Light Field Distortion (Wetzstein et al. ICCV2011)

Projection mapping
	Hirao?
\end{comment}

\vspace{-0.35cm}
\section{Refraction removal algorithm}
\vspace{-0.3cm}
\label{sec:algorithm}
In this section, we first introduce conventional ways to make refraction-free image (\subsecref{conventional}),
and then, introduce proposed method based on light field imaging (\subsecref{proposed}).

\vspace{-0.3cm}
\subsection{Virtual focus and lens distortion for approximation}
\begin{figure}[t]
	\begin{center}
		\includegraphics[width=7.0cm]{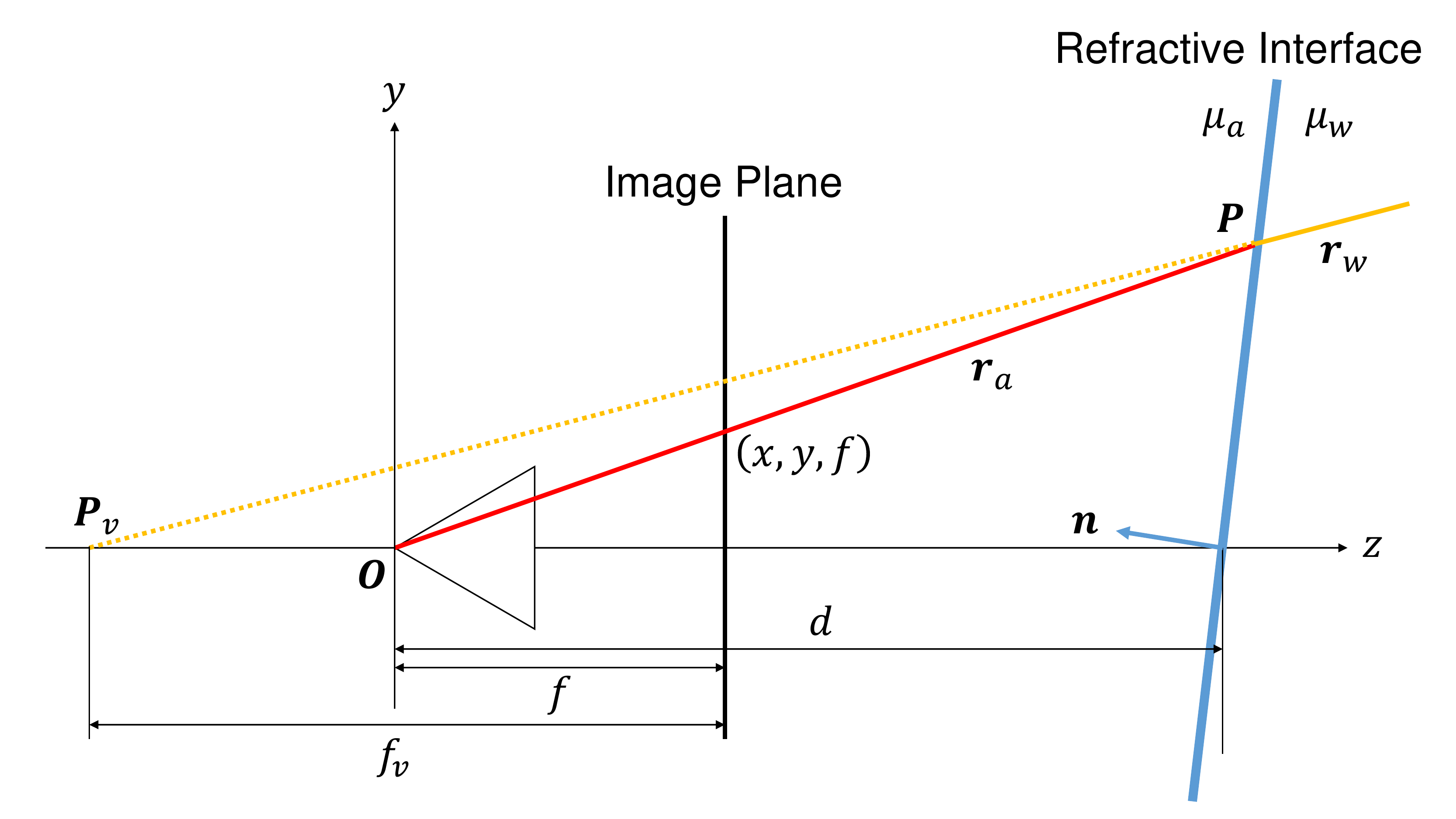}
		\vspace{-0.5cm}
		\caption{Illustration of approximation-based algorithm.}
		\label{fig:conv}
	\end{center}
	\vspace{-.9cm}
\end{figure}

\vspace{-0.1cm}
\label{ssec:conventional}
Let a camera coordinate system be a pinhole model with focal length $f$
% at origin point 
(\figref{conv}).
A planar refractive surface is placed in front of the camera with depth $d$ and normal ${\bm n}$.
Both sides of the refractive surface are filled with transparent medium with different refractive indices $\mu_a$ and $\mu_w$.
%In the figure, only  depth ($z$) and vertical ($y$) dimension are illustrated, 
%and horizontal ($x$) dimension is omitted for simplicity.
%since incident light and refracted light are on the same plane according to Snell's law.

When a light ray ${\bm r}_a$ is observed at $(x, y, f)$ on the image plane, 
the ray is backtracked to ${\bm P}$ on the refractive surface.
%, calculated as below:
%\begin{equation}
%	\bm{P} = \frac{\bm{n}\cdot[0\:0\:d]^T}{\bm{n}\cdot{\bm r}_a}{\bm r}_a
%	\label{eq:intersection}
%\end{equation}
As the ray is refracted on the refractive surface, the original direction ${\bm r}_w$ of the ray can be calculated based on Snell's law.
%As the ray is refracted on the refractive surface, the original direction ${\bm r}_w$ of the ray is:
%\begin{equation}
%	{\bm r}_w = \frac{1}{\mu}\left[{\bm r}_a - \left\{({\bm r}_a\cdot{\bm n})+\sqrt{\mu^2-1+({\bm r}_a\cdot{\bm n})^2}\right\}{\bm n}\right]
%	\label{eq:refraction}
%\end{equation}
%where $\mu = \mu_w/\mu_a$.
If the original ray did not refracted, it intersects with optical axis at ${\bm P}_v = {\bm P} + t{\bm r}_w$ with appropriate coefficient $t$.
Then we can define virtual focal length $f_v = f + |P_{vz}|$, as the system becomes central projection model.

However, since $f_v$ depends on ${\bm r}_a=[x, y, f]^T$, it varies with location on the image 
(That is why \cite{Kawahara:ICCVW2013} introduced pixel-wise varifocal camera model).
Several approximation-based approaches apply lens distortion model to convert 
original image 
into central projection image~\cite{Bleier:ISPRS2017,Kawasaki:WACV2012,Ichimaru:3DV2018}
by using $f_v$ and distortion coefficients, which are estimated by minimizing re-projection error.
%because differences of $f_v$ are observed as distortion on the image.
%In practice, $f_v$ and distortion coefficients are computed to minimize re-projection error.
Usually, radial and tangential distortion model are used to represent  the lens 
distortion, however, they can only approximate refraction at certain specific depth.
Thus, approximation error increases as target depth varies from predefined  depth~\cite{Ichimaru:3DV2018}, 
%to remove the distortion, represented with 4 or 5 parameters,
%which cannot completely explain refraction.
%Thus approximation error increases as target depth varies from calibrated depth.
%The characteristic of this depth-dependent error was analyzed in \cite{Ichimaru:3DV2018}, 
%to find out that the error rapidly increases when the target depth is closer than calibrated depth.
%It means close-range measurement is difficult with approximation-based approach.
especially, when the refractive interface is slanted, approximation error is biased to specific direction, which leads to severe error.

\vspace{-0.3cm}
\subsection{Ray selection from light field image}
\vspace{-0.1cm}
\label{ssec:proposed}

\begin{figure}[t]
	\begin{center}
		\includegraphics[width=7.0cm]{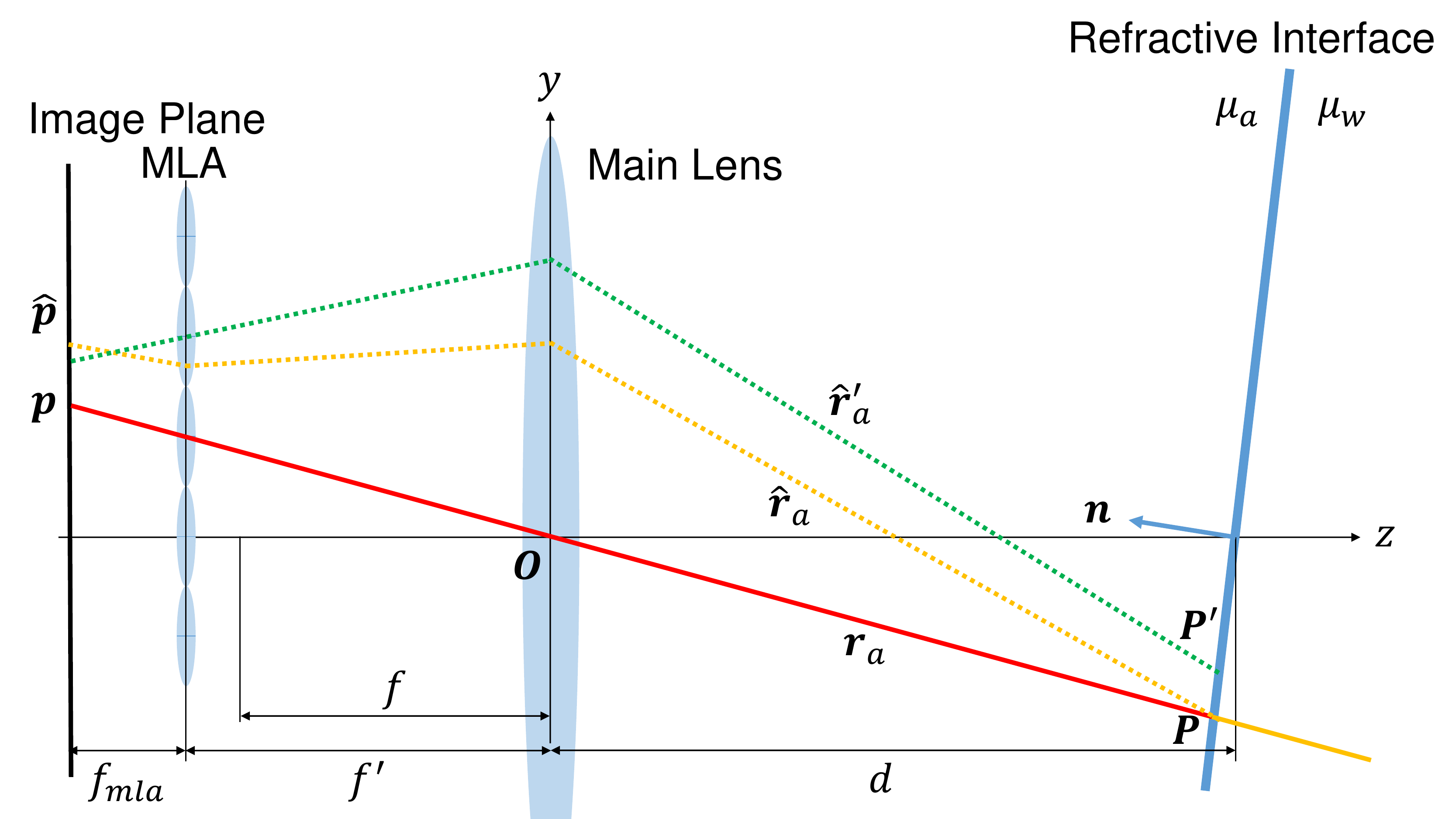}
		\vspace{-0.5cm}
		\caption{Illustration of proposed algorithm. Orange line represents necessary ray, and green line represents selected ray. }
		\label{fig:prop}
	\end{center}
	\vspace{-.9cm}
\end{figure}

To overcome the conventional distortion-free image synthesis technique based on 
approximation approach, we use light field image.
Let a light field camera consist of a single large lens (main lens), 
micro lenses aligned on a plane (micro lens array; MLA)
and image plane behind the MLA plane, as shown in \figref{prop}.
Main lens, MLA and image plane are parallel and all lenses are assumed to follow 
thin lens model. In the figure,
$f$ is a focal length of main lens, $f_{mla}$ is that of MLA and $f'$ is a distance between main lens and MLA.
Note that focal length $f$ is not equal to flange back $f'$ unlike pinhole camera model,
however, a distance between MLA and image plane is equal to $f_{mla}$.
The system represents physical light field camera model, instead of two-plane model.
%The rest is the same as \subsecref{conventional}.

When a light ray ${\bm r}_a$ is observed at ${\bm p}$ on the image plane, 
the ray is backtracked to ${\bm P}$ on the refractive surface.
%, calculated as (\ref{eq:intersection}).
If there were no refraction, the ray comes along straight line, 
however, it is refracted and reaches to another location $\hat{\bm p}$.
Therefore, by moving a color of $\hat{\bm p}$ onto location ${\bm p}$,
we can make refraction-free image.

In terms of calculation of $\hat{\bm p}$, we use ray tracing technique with thin 
lens model. %equation.
We can analytically compute which micro lens and which pixel the ray reaches.
However information of the ray is often lost because not all ray passes center of the micro lens.
Thus, instead of using the computed $\hat{\bm p}$, we select the closest ray from 
neighbor rays which passes through the center of the micro lens.
A distance between necessary ray and selected ray is defined as below:
\begin{equation}
	D=|{\bm P'}-{\bm P}|+\lambda\cos^{-1}(\hat{{\bm r}}_a'\cdot\hat{{\bm r}}_a),
\end{equation}
where $\hat{{\bm r}}_a'$ is direction of selected ray, $\hat{{\bm r}}_a$ is direction of necessary ray, 
${\bm P'}$ is intersection between $\hat{{\bm r}}_a'$ and the refractive 
surface and $\lambda$ is a weight. %balancing coefficient.
Once a ray which minimizes $D$ is found, ${\bm p}$ is assigned a color of a pixel the ray reached.

In practice, since it is almost impossible to calibrate $f$, it is assumed to be equal to $f'$.
The second term of $D$ always becomes zero based on %(we can derive the fact with 
simple calculation, thus, %which means 
only the distance from the intersections is minimized.
%So we did not set specific value to $\lambda$ in the implementation.

\vspace{-0.35cm}
\section{Implementation}
\vspace{-0.3cm}
\label{sec:impl}
\subsection{Refraction-free image synthesis}
\vspace{-0.1cm}
\label{ssec:removal}
To achieve efficient computation, we divided the synthesis process into two parts, such as light path computation and pixel warping part,
because light path computation is just once required through the entire process.
%once pixel warping map is generated by light path computation part, 
%we can just repeat applying pixel warping as long as the camera setting is the same.
In light path computation part, we analytically compute light path for each micro lens to find pixel-to-pixel correspondences, 
unless necessary light path goes to the outside the aperture.
%, based on the algorithm mentioned in \subsecref{proposed}.
We get locations of all micro lenses manually.
Once pixel-to-pixel correspondences are obtained, we apply weighted average and super-sampling for better image quality.
For weighted average calculation, we extract several rays in ascending order of 
$D$ and average their color intensities according to reciprocal number of each $D$.
For super-sampling, we increase the number of pixel-to-pixel correspondences by computing subpixel rays.
The number of rays for average and the super-sampling ratio are changed in our implementation.
We finally obtain a pixel warping map based on weight for average and 
super-sampling for each pixel.

In pixel warping part, we simply use the pre-computed pixel warping map to light field image
and final image with specified resolution is obtained by bilinear interpolation.

\vspace{-0.3cm}
\subsection{Underwater stereo using refraction-free image}
\vspace{-0.1cm}
\label{ssec:stereo}
%We also describe scheme of underwater passive stereo using refraction-free images made by our algorithm.
%Schematic of the scheme is shown in \figref{stereo}.
%with fixed relative transformation, 
%and they are set in the planar waterproof housing filled with air, to be submerged into the liquid.

\begin{figure}[t]
	\begin{center}
		\includegraphics[width=8.0cm]{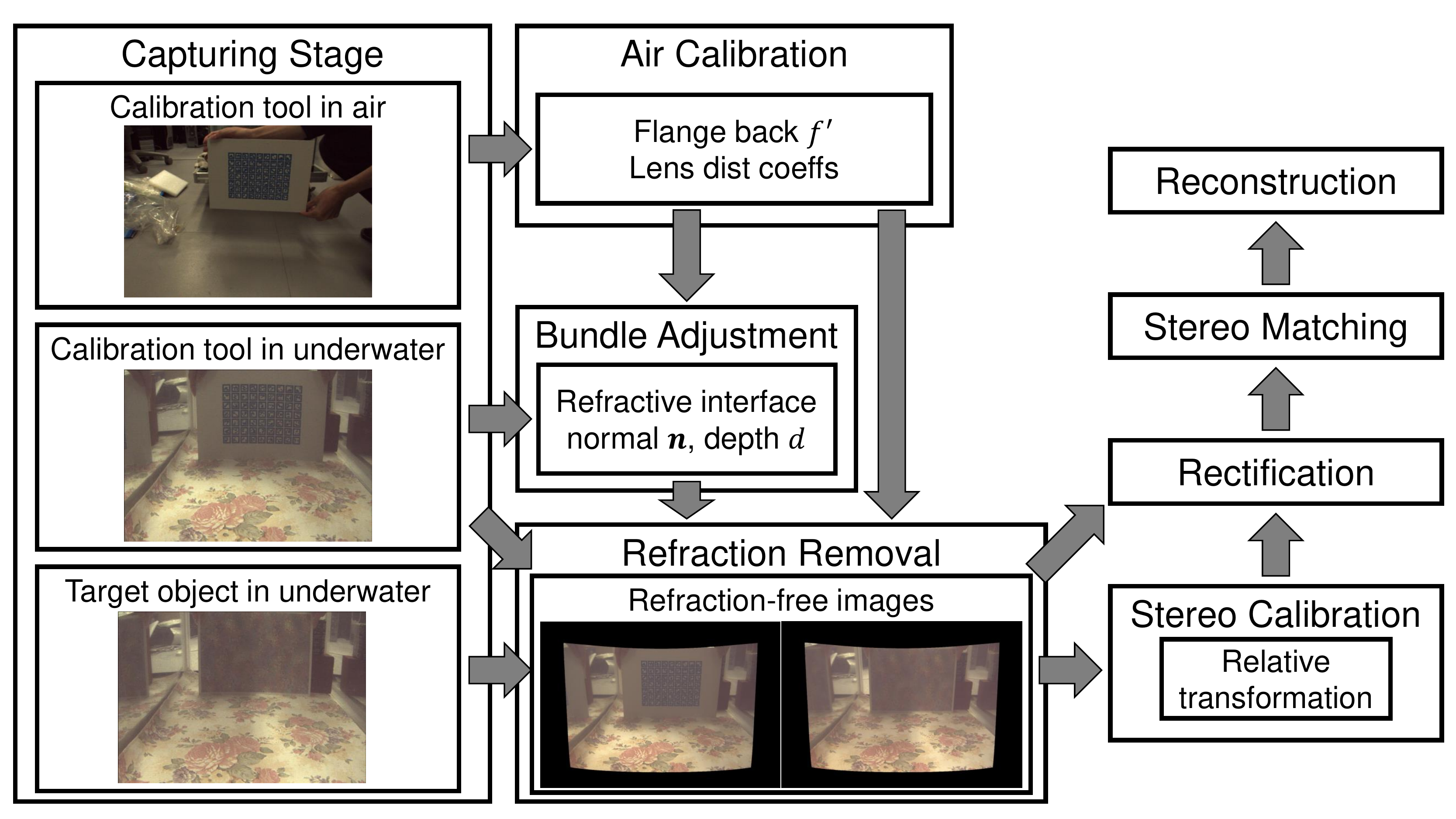}
		\vspace{-0.5cm}
		\caption{Schematic of refraction-free stereo.}
		\label{fig:stereo}
	\end{center}
	\vspace{-.9cm}
\end{figure}

Two light field cameras are used to conduct underwater stereo.
Overview of the process is shown in \figref{stereo}.
First, calibration tools (\eg, chess patterned board) are captured both in the air and the liquid
and 2D locations of feature points are obtained.
Second, real flange back $f'$ and lens distortion coefficients of the cameras are calibrated with images captured in the air.
Cameras' real lens distortion effect is removed from all images at this stage.
Calibration of relative transformation of the cameras is not necessary here, thus air images can be captured separately.
Third, $d$ and ${\bm n}$ of the refractive interface are calibrated by bundle 
adjustment using the method of \cite{Agrawal:CVPR2012}.
Then, refraction-free images are synthesized using proposed algorithm with obtained parameters $f'$, $d$, and ${\bm n}$.
Using the image set, relative position and orientation between camera pair are calibrated with the refraction-free images and finally stereo method is applied.
As for the stereo algorithm, OpenCV with NCC based matching cost is applied.

%Of course, proposed algorithm can be used not only for passive stereo, 
%but also for active or single-view methods as long as refraction effects of other optical devices are solved.

\vspace{-0.35cm}
\section{Experiment}
\vspace{-0.3cm}
\label{sec:experiment}
\subsection{Evaluation using planar object}
\begin{figure}[t]
	\begin{center}
		\includegraphics[width=5.0cm]{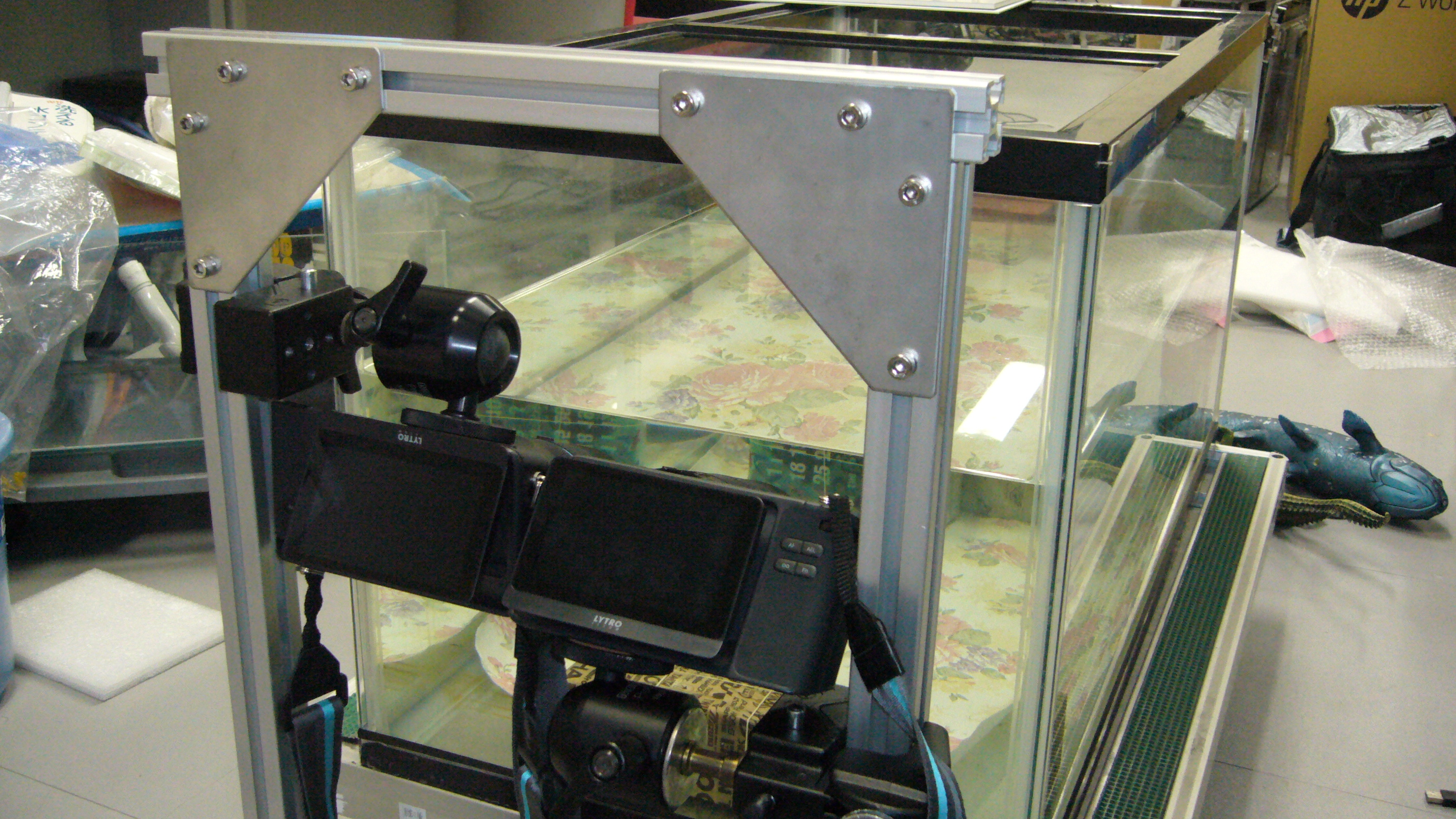}
		\vspace{-0.4cm}
		\caption{Appearance of the experimental setup.}
		\label{fig:setup}
	\end{center}
	\vspace{-0.8cm}
\end{figure}

%In this section, we describe the experiment for evaluating proposed method.
For the experiment, two light field cameras, Lytro Illum, 
and water tank of $90\times45\times45cm$ dimensions %size %as refractive interface 
are used (\figref{setup}).
Two cameras are set outside the water tank, and the water tank is filled with clear water.
%One camera is upside down to make the optical axes as close as possible for close-range capturing.
We intentionally slanted the cameras (it looks as if the refractive interface is slanted for the cameras), 
to make a strong distortion with approximation methods, %for conventional approach, 
which is expected to be solved by our technique.
%examine how proposed algorithm is superior under such irregular situation.
%
We captured calibration board at two distances, such as far 
%position from the cameras 
and near from the cameras
to see the tolerance of proposed algorithm against depth-dependent error.
We applied both proposed algorithm as well as approximation-based 
algorithm~\cite{Ferreira:PRIA2005} to synthesize distortion-free images.
%for calibration following to the scheme mentioned in \subsecref{stereo} to the captured images.

\begin{figure}[t]
	\begin{minipage}{0.49\hsize}
		\begin{center}
			\includegraphics[width=4cm]{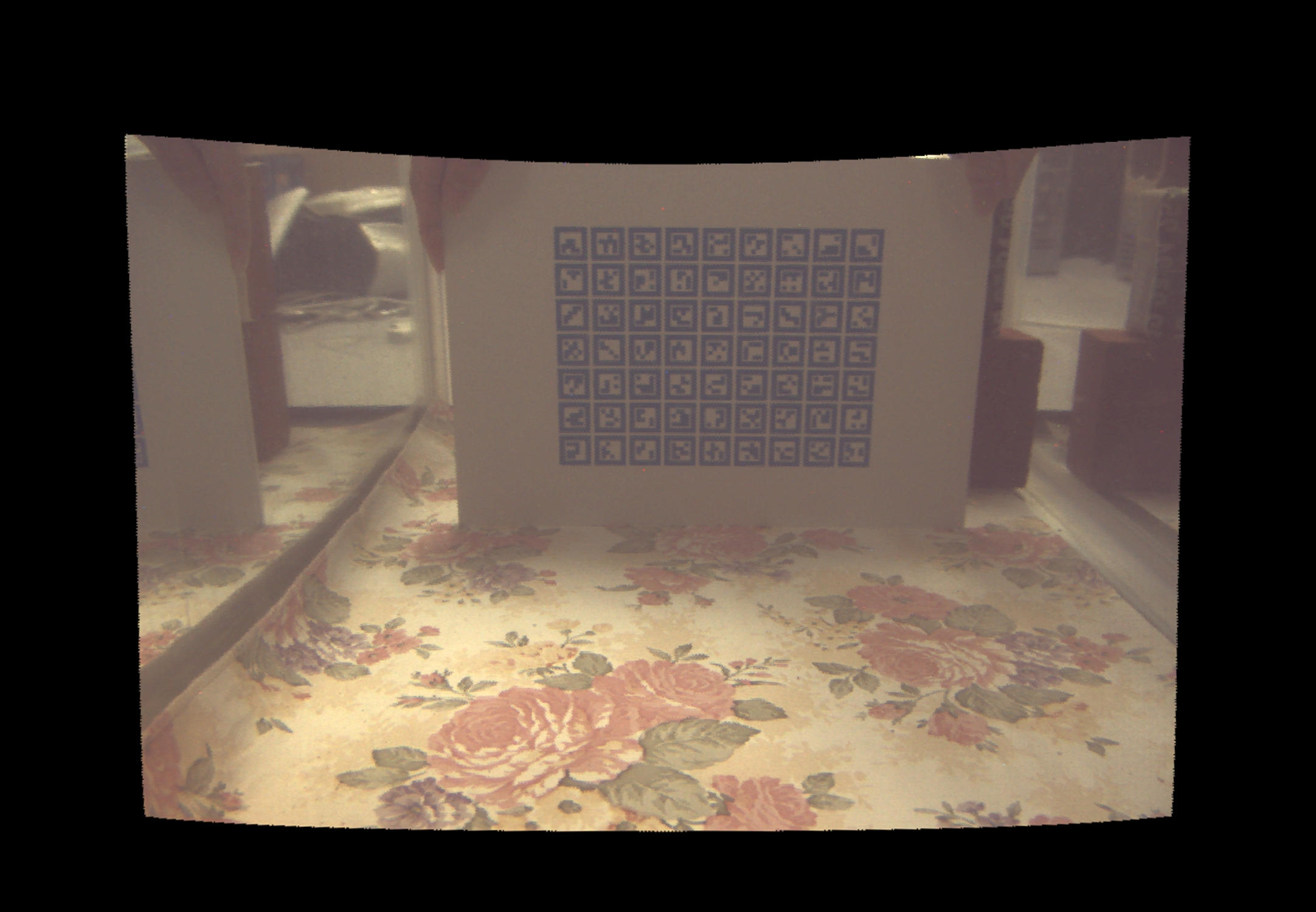}
		\end{center}
	\end{minipage}
	\begin{minipage}{0.49\hsize}
		\begin{center}
			\includegraphics[width=4cm]{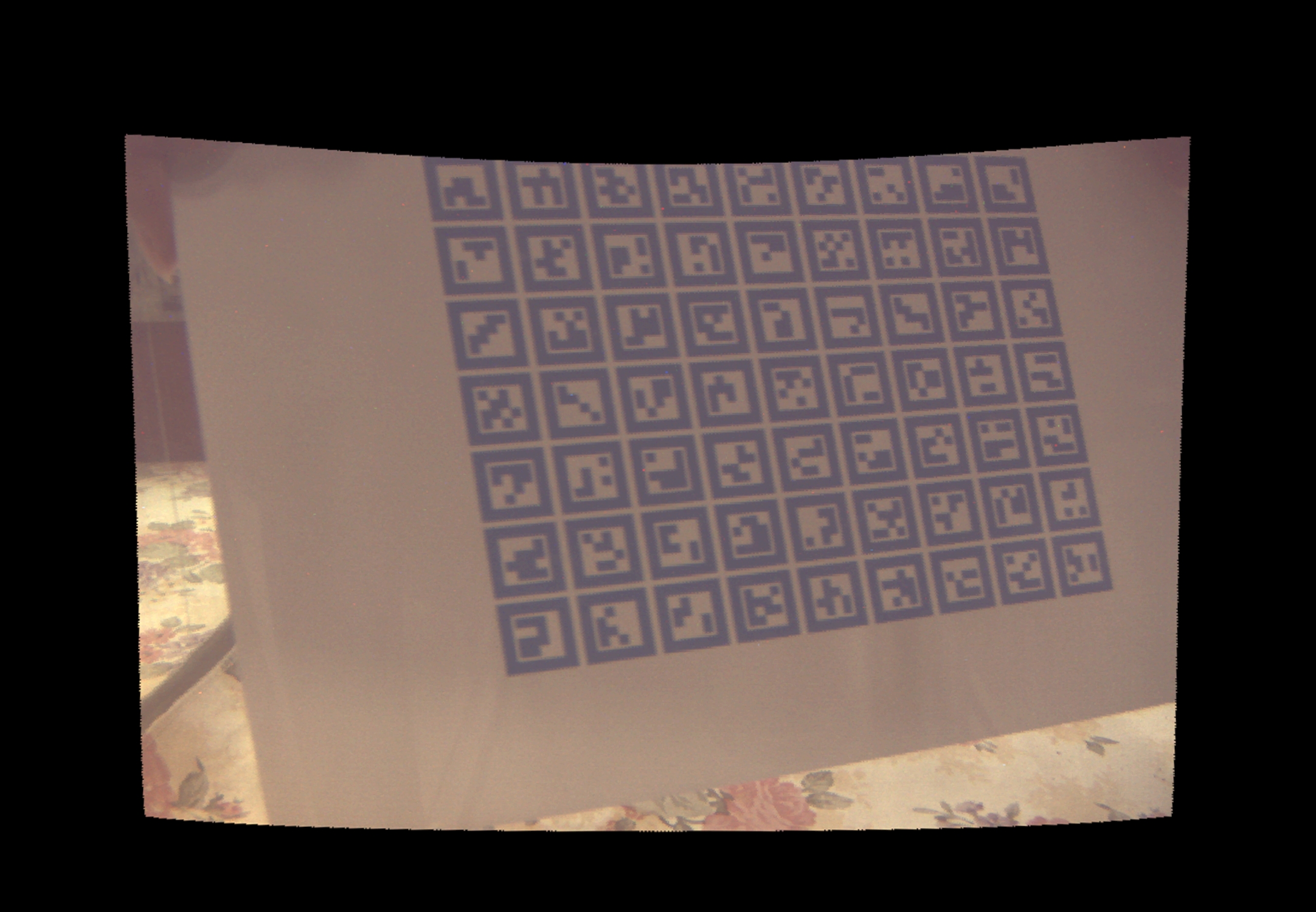}
		\end{center}
	\end{minipage}
	\vspace{-0.2cm}
	\caption{Examples of refraction-free images with our method.}
	\label{fig:reffree}
	\vspace{-0.5cm}
\end{figure}

For evaluation, we captured the textured planar board at far, medium and near 
from the camera.
%for shape reconstruction.
We tested with following four different conditions, such as (a) approximation-based algorithm~\cite{Ferreira:PRIA2005} with far calibration board, 
(b) approximation-based algorithm~\cite{Ferreira:PRIA2005} with near calibration board, (c)
proposed algorithm and (d) proposed algorithm with distortion removal based on 
~\cite{Ferreira:PRIA2005}. %approximation-based algorithm with both calibration board.
We set number of weighted averaging pixels to 32, and super-sampling ratio to 2.
Examples of refraction-free images are shown in \figref{reffree}.
%It took 8.9 seconds for pre-computation of pixel warping map, and 53 milliseconds for pixel warping and other image processing 
%with Intel Xeon E5640 CPU, showing lightweight computation cost of proposed algorithm.
Image synthesis takes 53 milliseconds for our algorithm
with Intel Xeon E5640 CPU, which is a realtime process.
%significantly faster than geometric approach.

\begin{figure}[t]
	\begin{center}
		\includegraphics[width=8.0cm]{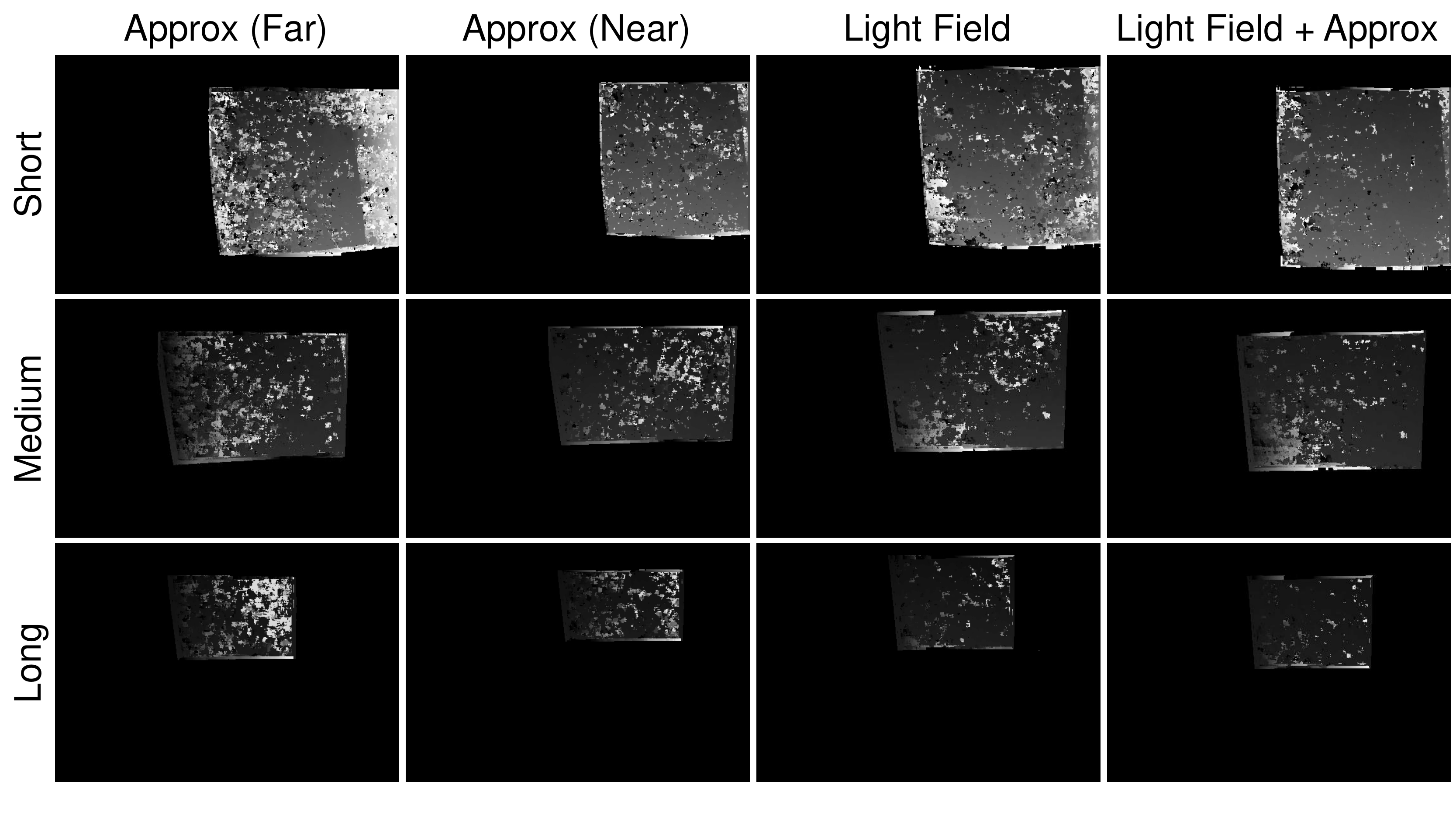}\\
		\vspace{-0.2cm}
(a) \hspace{1.5cm} (b) \hspace{1.5cm} (c) \hspace{1.5cm} (d)\\
		\vspace{-0.3cm}
		\caption{Results of stereo matching with four methods. 
(a) and (b): approximation-based algorithm~\cite{Ferreira:PRIA2005}. (c) and (d):
proposed algorithm. Noises are decreased in our algorithm.}
%``Approx'' means approximation-based algorithm, and ``Light Field'' means proposed algorithm.}
		\label{fig:disparity}
	\end{center}
	\vspace{-0.8cm}
\end{figure}

\begin{figure}[t]
	\begin{center}
		\includegraphics[width=7.0cm]{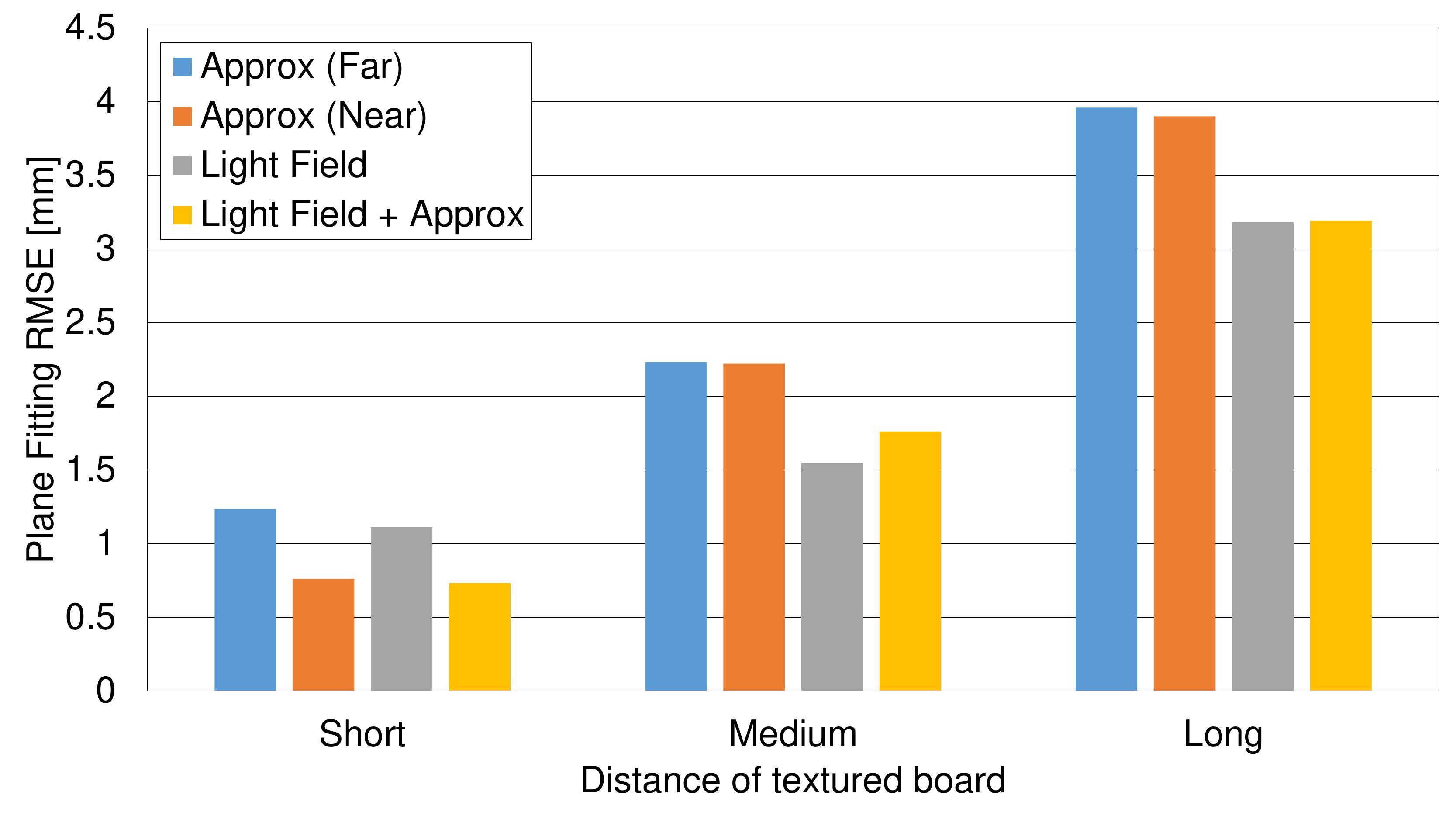}
		\vspace{-0.5cm}
		\caption{Results of plane fitting on textured board.}
		\label{fig:graph}
	\end{center}
	\vspace{-0.8cm}
\end{figure}

After synthesizing refraction-free images using each method, we 
rectified the images and applied NCC based stereo.
% matching to reconstruct the shape of the board.
Results of disparity maps are shown in \figref{disparity}.
We can observe that the approximation-based algorithms~\cite{Ferreira:PRIA2005} 
produces moderate results when the target depth is close to the depth of the calibrated tool, 
however, the results 
are getting worse, if the target depth is far from the depth of the calibration tool.
On the other hand, proposed algorithms produce better results regardless of 
depth of the target and the calibration board.
\figref{graph} shows quantitative evaluation results by plane fitting on 
reconstructed board after outlier removal, clearly showing 
%We can confirm 
that the proposed algorithm outperforms the approximation-based 
techniques quantitatively.

\vspace{-0.3cm}
\subsection{Reconstruction of arbitrary shape objects}
\vspace{-0.1cm}

\begin{figure}[t]
	\begin{minipage}{0.32\hsize}
		\begin{center}
			\includegraphics[width=3.0cm]{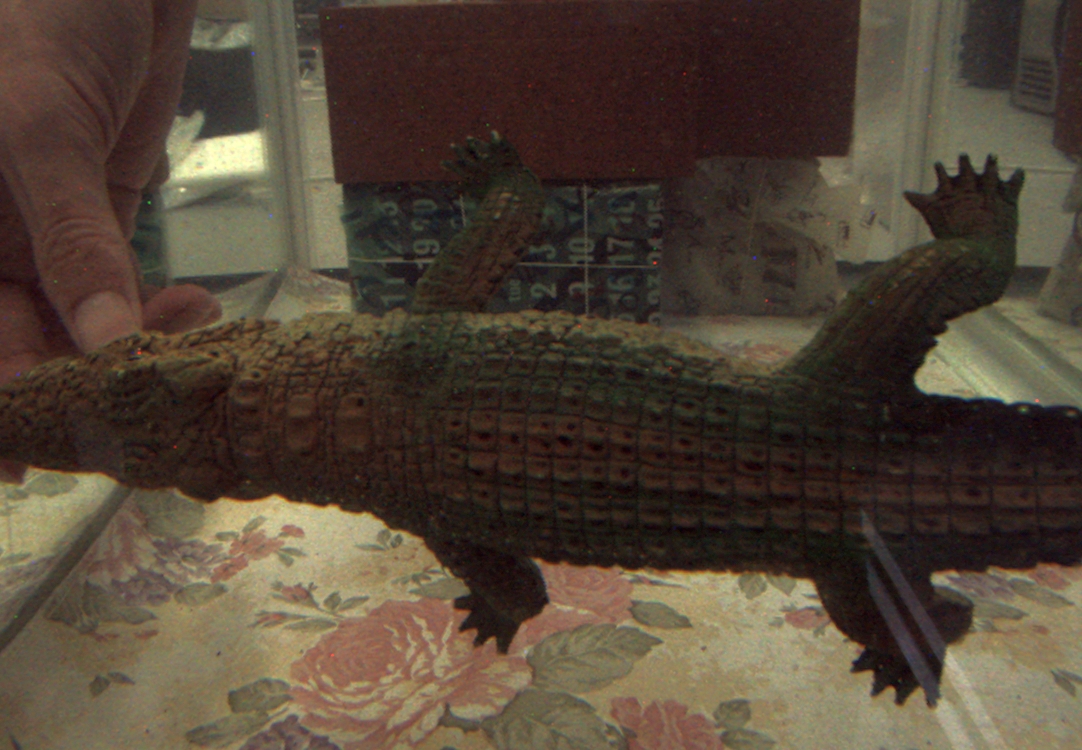}
		\end{center}
	\end{minipage}
	\begin{minipage}{0.32\hsize}
		\begin{center}
			\includegraphics[width=3.0cm]{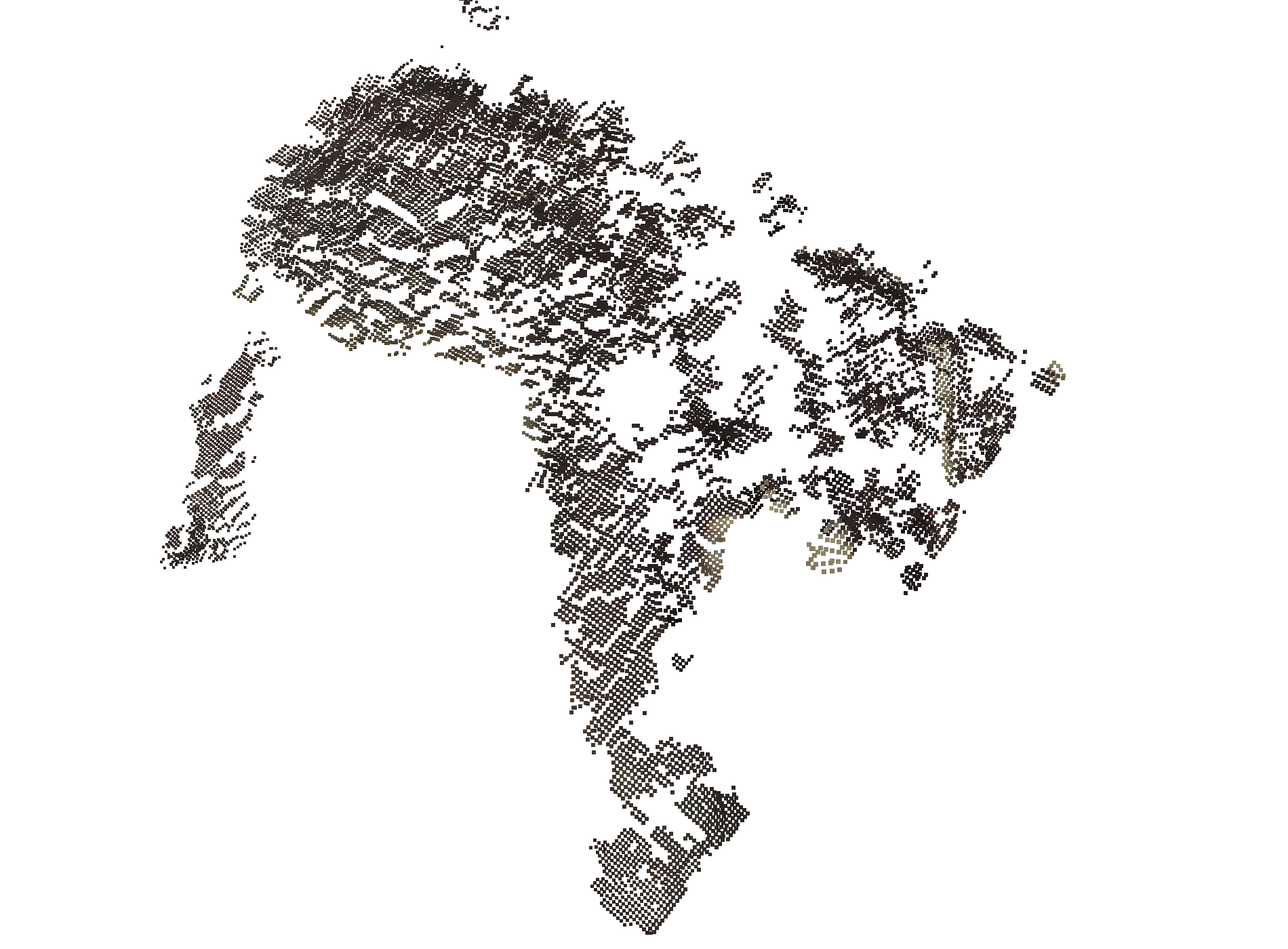}
		\end{center}
	\end{minipage}
	\begin{minipage}{0.32\hsize}
		\begin{center}
			\includegraphics[width=3.0cm]{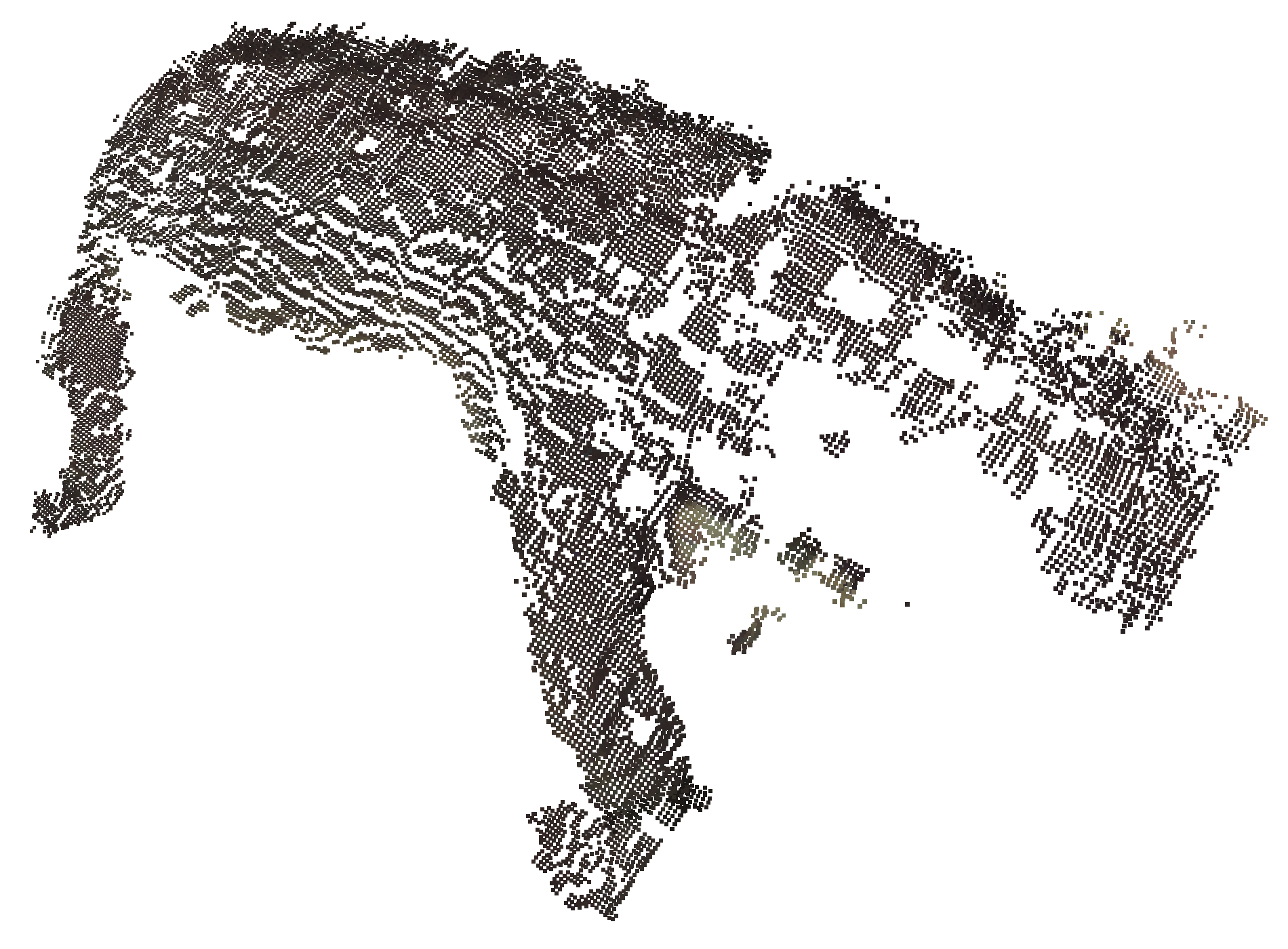}
		\end{center}
	\end{minipage}
	\begin{minipage}{0.32\hsize}
		\begin{center}
			\includegraphics[width=3.0cm]{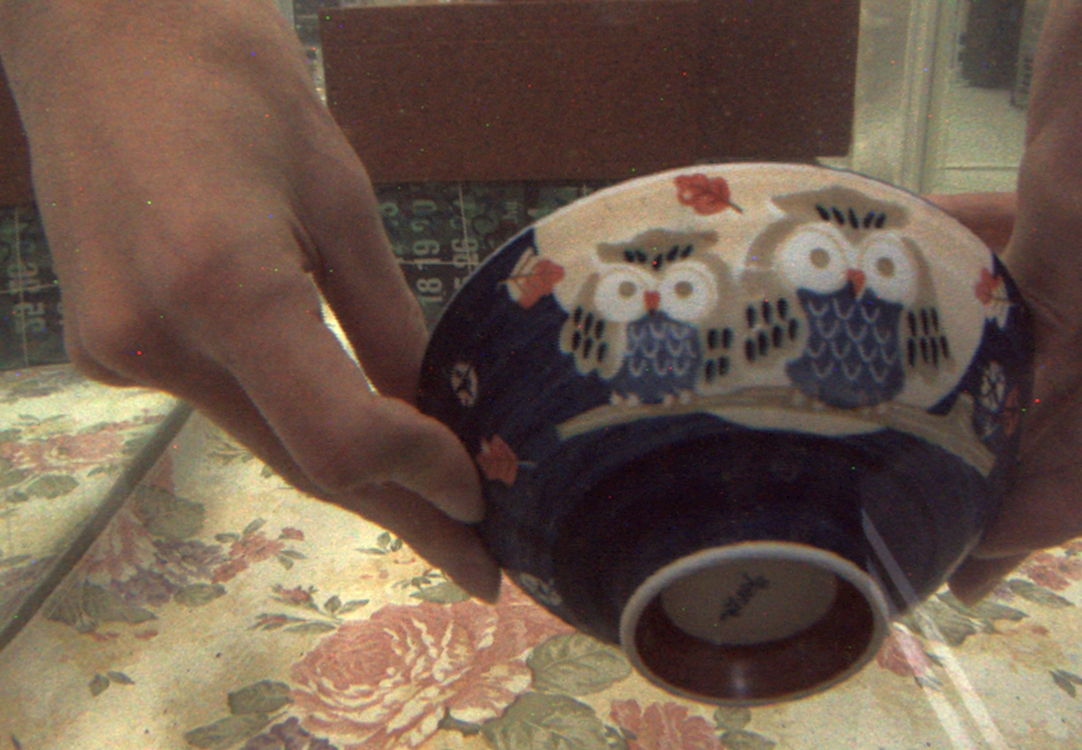}
		\end{center}
	\end{minipage}
	\begin{minipage}{0.32\hsize}
		\begin{center}
			\includegraphics[width=2.0cm]{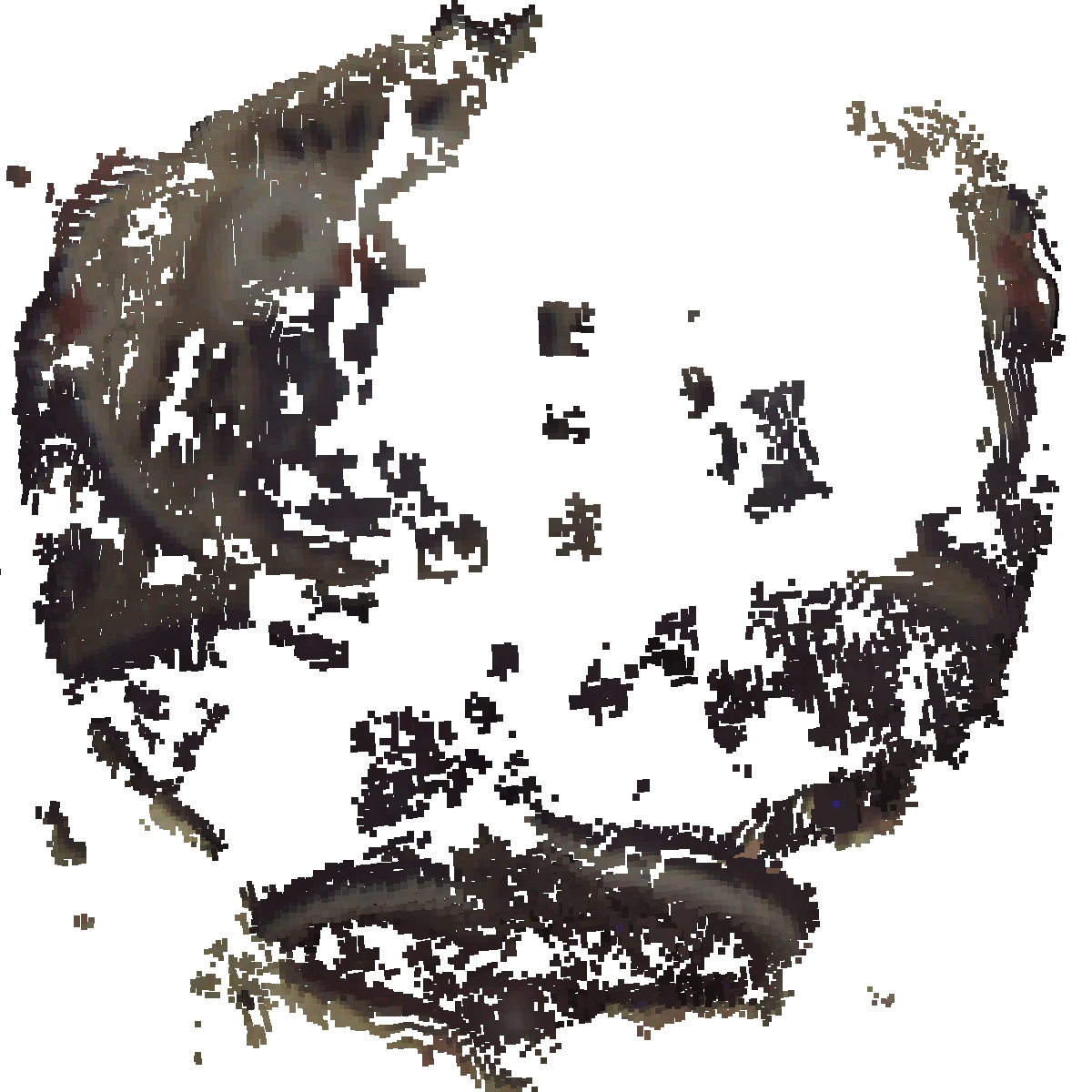}
		\end{center}
	\end{minipage}
	\begin{minipage}{0.32\hsize}
		\begin{center}
			\includegraphics[width=2.0cm]{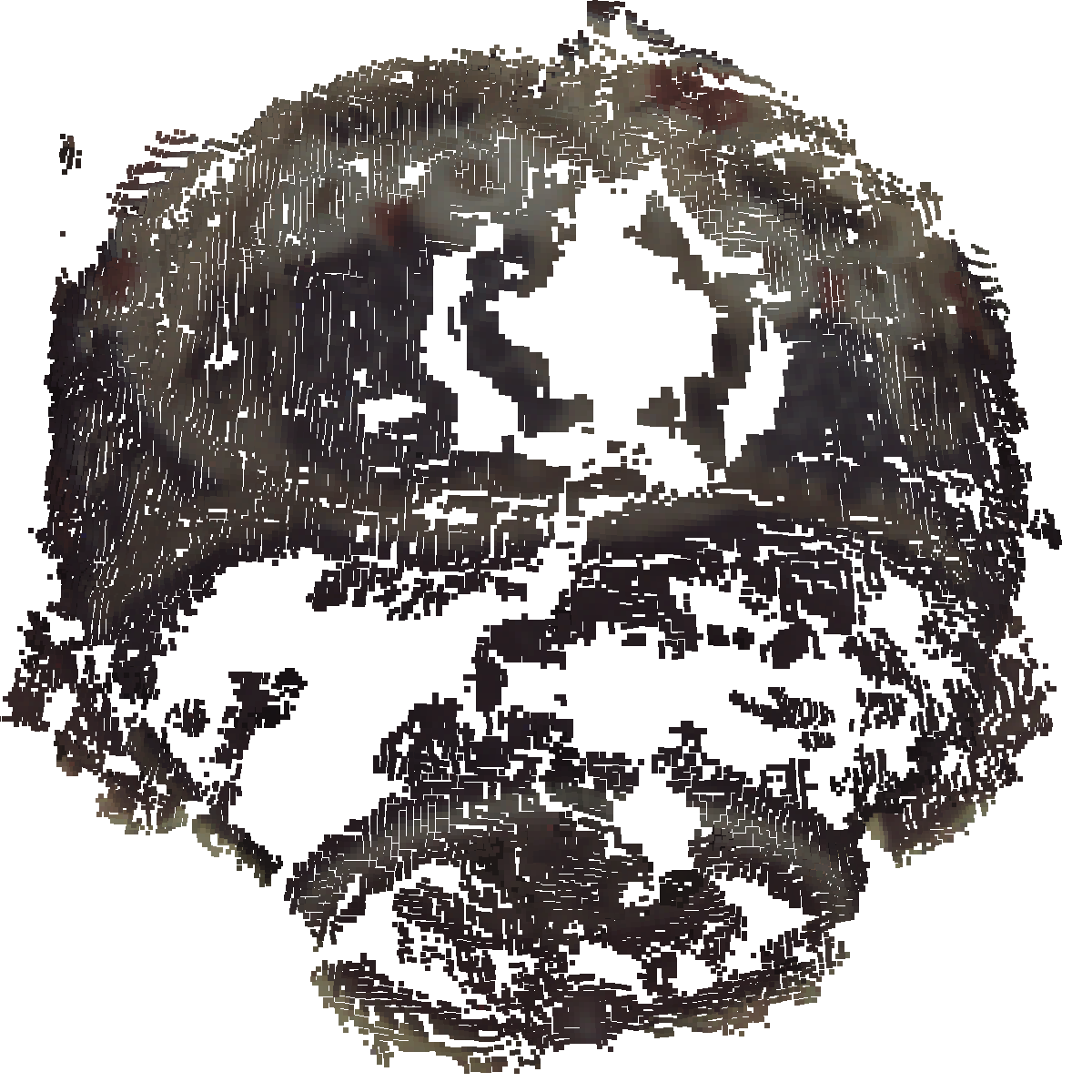}
		\end{center}
	\end{minipage}
	\begin{center}
		\vspace{-0.5cm}
		\caption{Qualitative comparison on reconstruction. {\bf Left: } Captured images. {\bf Center: } Results of approximation-based algorithm~\cite{Ferreira:PRIA2005}. {\bf Right: } Results of proposed algorithm. }
		\label{fig:others}
	\end{center}
	\vspace{-1.0cm}
\end{figure}

We tested our method using several objects with complicated
texture, such as a crocodile figurine and a ceramic bowl.
In this experiment, depth of the calibration board is placed farther than both target objects.
Reconstruction results for qualitative evaluation are shown in \figref{others}.
We also apply conventional approximation-based algorithm~\cite{Ferreira:PRIA2005} for comparison.
We can confirm that the shapes of the approximation-based algorithm is unstable 
because of the mismatch of the depth between reconstruction and calibration,
whereas our technique achieved stable reconstruction for both objects.

\vspace{-0.35cm}
\section{Limitation}
\begin{figure}[t]
	\begin{center}
		\includegraphics[width=6.0cm]{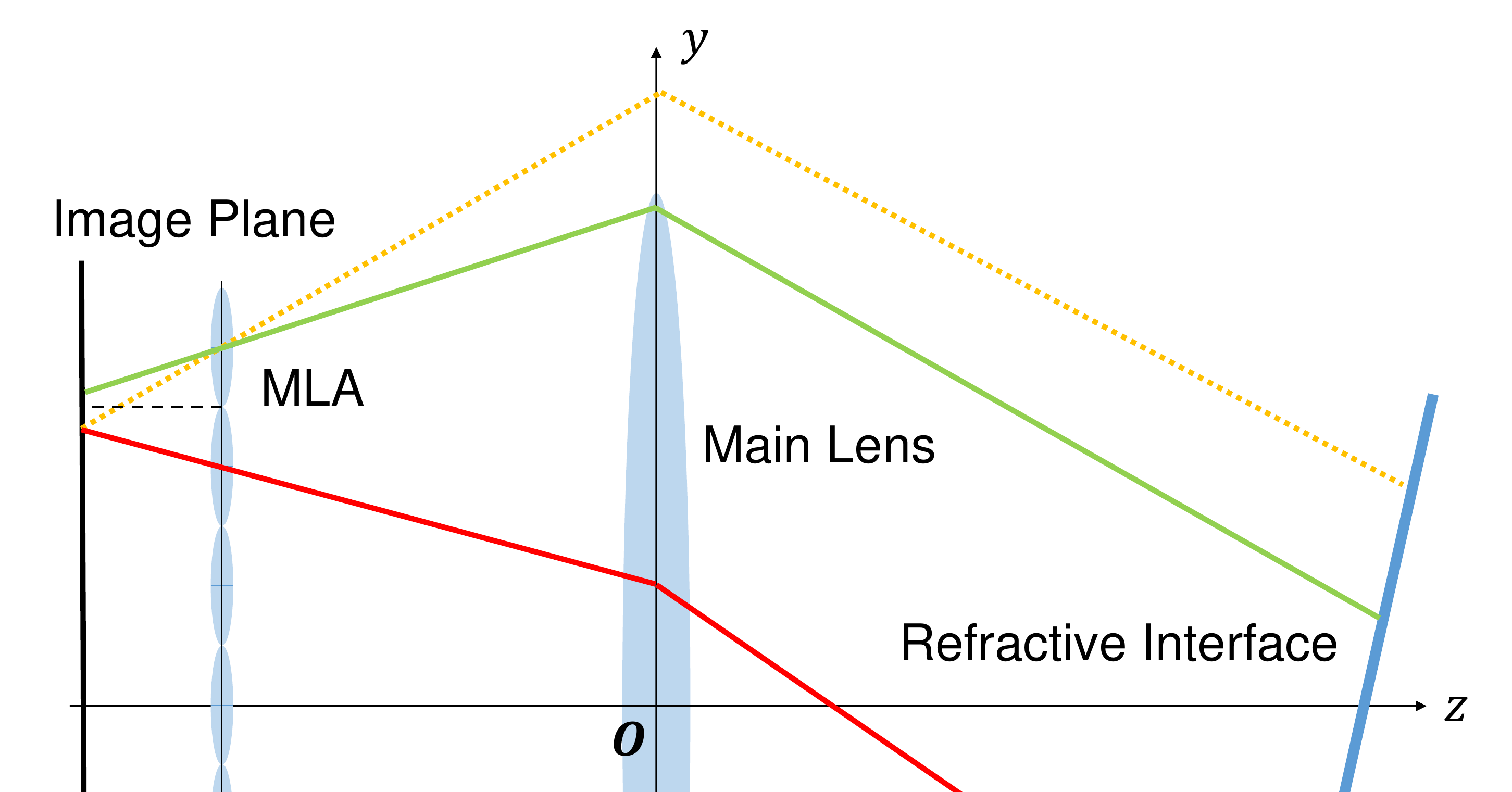}
		\vspace{-0.3cm}
		\caption{Illustration of the limitation and compensation by weighted averaging. Since red line ray is observed instead of orange line going outside the aperture, we collect neighbor green line rays to synthesize orange line ray.}
		\label{fig:limit}
	\end{center}
%	\vspace{-0.8cm}
%\end{figure}
%
\vspace{-0.2cm}
%
%\begin{figure}[t]
	\begin{minipage}{0.32\hsize}
		\begin{center}
			\includegraphics[width=2.8cm]{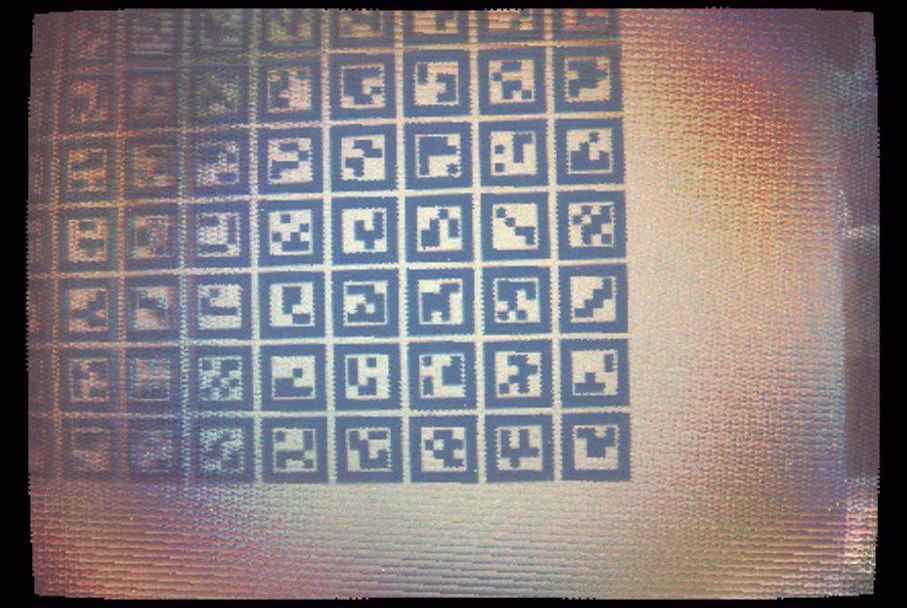}
		\end{center}
	\end{minipage}
	\begin{minipage}{0.32\hsize}
		\begin{center}
			\includegraphics[width=2.8cm]{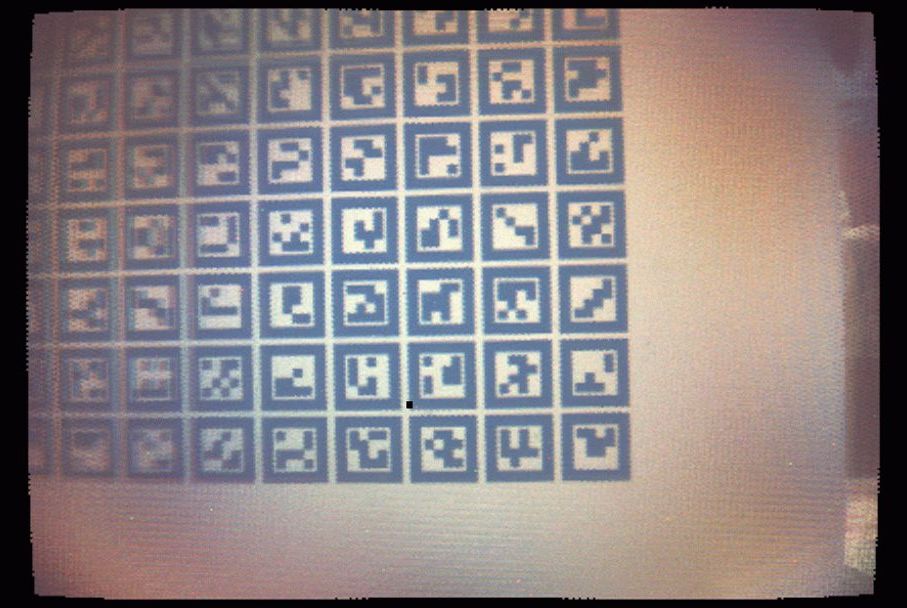}
		\end{center}
	\end{minipage}
	\begin{minipage}{0.32\hsize}
		\begin{center}
			\includegraphics[width=2.8cm]{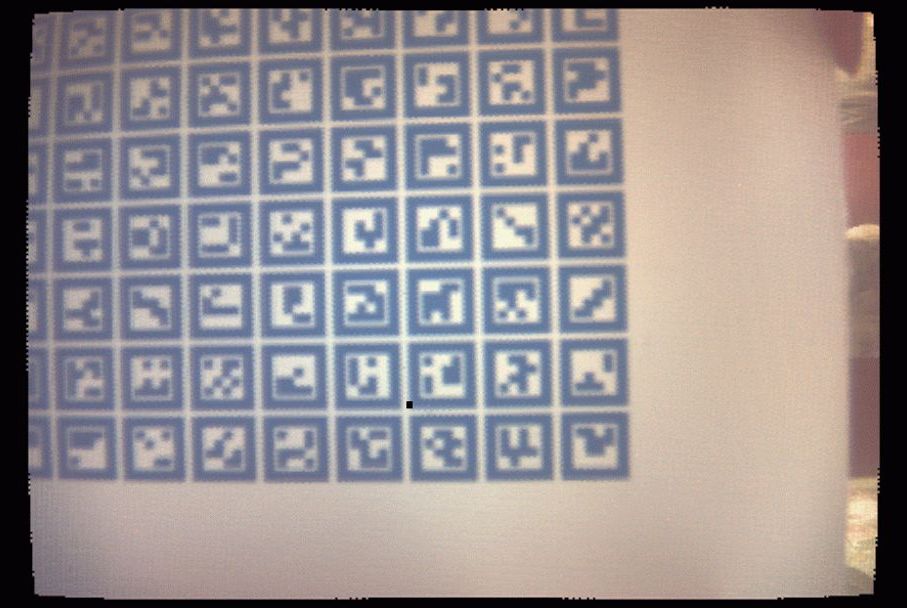}
		\end{center}
	\end{minipage}
	\begin{center}
		\vspace{-0.5cm}
		\caption{Comparison of synthesized images with different numbers of averaged rays (1, 8, 64, from left to right).}
		\label{fig:ave_comp}
	\end{center}
	\vspace{-1.0cm}
\end{figure}

\vspace{-0.3cm}
\label{sec:limitation}
In the experiment, we achieved higher accuracy in our algorithm than conventional algorithms.
However, in practice, approximation-based algorithm can produce 
similar/sometimes even better quality 
than our technique, 
if refractive interface is precisely orthogonal to the camera axis.
We consider this is because our algorithm can produce perfect refraction-free image in theory, 
however, due to the limitation of the aperture size,
image quality can be degraded by the following reason.
%If necessary ray goes outside the aperture, we cannot retrieve color for the pixel.
According to our calculation, possible field-of-view for perfect refraction-free image is 30 degrees, 
while original field-of-view is about 60 degrees for Lytro Illum.
% is used to capture images.
To compensate the problem, we introduce 
%such unavailability of wide view by 
a weighted average of nearby pixels in our method (\figref{limit}), 
which leads a defocus effect producing blur in final image
%since the image get close to the original F-number 
by increasing the number of averaging rays (\figref{ave_comp}).
Using wider field-of-view for the lens is one promising solution.

%However, approximation-based approach can only model simple refraction effect, observed as radial or tangential distortion.
%In this paper, we just slanted the refractive interface, 
%but the true capability of proposed algorithm is utilized when the interface is thick or complicated shape, 
%as long as geometrically correct light ray tracing is possible.

\vspace{-0.35cm}
\section{Conclusion}
\vspace{-0.3cm}
\label{sec:conclusion}
In this paper, we propose an algorithm to synthesize refraction-free image using light field camera.
Proposed algorithm enabled geometrically correct image synthesis for any kind of 
refraction 
and performed better 3D reconstruction based on stereo than previous 
approximation-based methods. %algorithm. % under irregular refraction.
Moreover, the proposed method is capable of fast-computation, \ie,
real-time process with common PC.
Although there is a limitation due to the limited size of aperture, 
our experiments show that our technique is mostly better than previous methods, 
especially under severe refraction cases.
%which are 
%efficiently compensated by weighted average interpolation.
%however it was acceptable in our experiment.
As our future work, we consider underwater single view depth computation is also possible using light field camera, as well as applying the proposed method to active stereo techniques.

\clearpage

\bibliographystyle{IEEEbib}
\bibliography{strings,refs}

\end{document}